\documentclass[
 reprint,
 amsmath,
 amssymb,
 aps,superscriptaddress
]{revtex4-2}

\usepackage{graphicx}% Include figure files
\usepackage{dcolumn}% Align table columns on decimal point
\usepackage{bm}% bold math
\usepackage{mathtools}
\usepackage{comment}
\usepackage{float}
\usepackage{subfig}
\usepackage{ragged2e}
\usepackage{tikz}
\DeclareCaptionJustification{justified}{\justifying}
\usepackage{placeins}

\newenvironment{captivy}[1]{%SETUP
  \begin{tikzpicture}[every node/.style={inner sep=0}]
    \node[anchor=south west,inner sep=0] (image) at (0,0) {#1};
    \begin{scope}[x={(image.south east)},y={(image.north west)}]
}%
{%TEARDOWN
        \end{scope}%
  \pgfresetboundingbox
  \path[use as bounding box] (image.south west) rectangle (image.north east);
  \end{tikzpicture}%
}

\newcommand*{\oversubcaption}[3]{%
  \draw (#1) node[fill=white,inner sep=0pt, opacity=0.2, above, yscale=1.1, xscale=1.1] {\phantom{(a)#2}};
  \draw (#1) node[inner sep=0pt, above]{%
    \subfloat[#2\label{#3}]{\phantom{(a)}}
  };
}

\newcommand{\addtuc}{School of Electrical and Computer Engineering,Technical University of Crete, Chania, Greece 73100}

\newcommand{\addcqt}{Centre for Quantum Technologies, National University of Singapore, 3 Science Drive 2, Singapore 117543}

\newcommand{\addaq}{AngelQ Quantum Computing, 531A Upper Cross Street \#04-95 Hong Lim Complex, Singapore 051531}

\begin{document}

\title{Unravelling quantum chaos using persistent homology}

\author{Harvey Cao}
\email{harvey.cao@u.nus.edu}
\affiliation{\addcqt}
\author{Daniel Leykam}
\email{daniel.leykam@gmail.com}
\affiliation{\addcqt}
\author{Dimitris G. Angelakis}
\email{dimitris.angelakis@gmail.com}
\affiliation{\addcqt}
\affiliation{\addtuc}
\affiliation{\addaq}

\date{\today}

\begin{abstract}
Topological data analysis is a powerful framework for extracting useful topological information from complex datasets. Recent work has shown its application for the dynamical analysis of classical dissipative systems through a topology-preserving embedding method that allows reconstructing dynamical attractors, the topologies of which can be used to identify chaotic behaviour. Open quantum systems can similarly exhibit non-trivial dynamics, but the existing toolkit for classification and quantification are still limited, particularly for experimental applications. In this work, we present a topological pipeline for characterizing quantum dynamics, which draws inspiration from the classical approach by using single quantum trajectory unravelings of the master equation to construct analogue `quantum attractors' and extracting their topology using persistent homology. We apply the method to a periodically modulated Kerr-nonlinear cavity to discriminate parameter regimes of regular and chaotic phase using limited measurements of the system. 
\end{abstract}

\maketitle

\section{\label{sec:intro}Introduction}

Classical nonlinear systems can exhibit chaotic behaviour if they possess extreme sensitivity to initial conditions and are topologically mixing. The interplay of these two properties leads to long-term unpredictability and seemingly random dynamics, despite being fully deterministic. When dissipative effects play a role, a complex geometrical structure can be observed in the phase space dynamics known as a `strange attractor'. Strange attractors differ from other phase space attractors in that two points that are initially nearby on the attractor will eventually evolve to become arbitrarily far apart. This divergence is used as a diagnosis of whether a particular system is chaotic, quantified in conventional methods by calculating the maximal Lyapunov exponent (LE), a measure of the average rate of divergence in phase space~\cite{ott_2002}.

When quantum effects are significant, chaotic behaviour at the wavefunction level is precluded due to the linearity of the Schrödinger equation \cite{Berry_1989}. Instead, the role of chaos in quantum mechanics is studied by identifying signatures of quantum systems which result in a classical counterpart that yields chaotic dynamics. Early efforts in quantifying quantum chaos focused on the spectral statistics of chaotic Hamiltonians, with level spacing statistics corresponding to that of random matrices when the classical limit is chaotic \cite{berry_level_1977, haake_classical_1987,kolovsky_quantum_2004}. There have also been various quantum generalisations of the classical Lyapunov exponent constructed in terms of quasiclassical phase-space dynamics, the stochastic Schrödinger equation and, more recently, out-of-time-order correlators (OTOCs) \cite{haake_lyapunov_1992, manko_lyapunov_2000, bhattacharya_2000, rozenbaum_2017}. Beyond addressing fundamental questions concerning the correspondence principle between quantum and classical limits \cite{Wang_2021}, there is interest in understanding the role of quantum chaos in the thermalization of quantum many-body systems and subsequent implications for quantum information and near-term quantum computation \cite{sahu_2022, Borgonovi_2016, joshi_2022, frahm_2004, boixo_2018}.

In comparison to the classical case, where the maximal LE can be used unanimously as an indicator of chaos, the toolset for quantifying chaos in open quantum systems remains limited. Furthermore, these methods typically involve quantifiers that are difficult to measure in real experimental settings. For example, calculating the recently proposed OTOC-based LEs requires the ability to precisely reverse the evolution of a system, which is extremely challenging \cite{Swingle_2016, Li_2017}.

Recently, there has been growing interest in topological data analysis (TDA) methods for approaching problems in physics where geometric structure and symmetries can be exploited \cite{carlsson_2020,murugan_2019,leykam_2022}. TDA is a powerful and robust framework that enables the extraction of useful and explainable topological information from complex high-dimensional datasets. One such application has been in the characterization of classical dynamical systems through counting the number of low-dimensional holes of reconstructed attractors \cite{maletic_2016, mittal_bifur_2017, tran_2019, templeman_2020}. This offers an alternative to methods based on calculating the maximal LE, which can be heavily affected by noise and finite data sizes \cite{wolf_1985, margaris_2009}.

The primary objective of this study is to see whether TDA-based methods for chaos detection which have already proven effective in the classical case can be applied to the harder-to-visualise dynamics of open quantum systems, where the attractor in phase space is no longer well-defined by trajectories as in the classical case. As single system trajectories are compared in the classically chaotic case to find exponential divergence, it seems sensible to adopt a quantum approach that does likewise. Thus, we propose an approach based on quantum trajectories that are obtained from unravelling methods. Specifically, we consider the Monte Carlo wavefunction method, which replaces the original master equation with an ensemble of stochastically evolving quantum trajectories, whose average recovers the full density operator evolution \cite{BRE02, Molmer_1993}. We follow one member of the ensemble and use the time-evolution of suitable observables to approximate a classical trajectory analogue. 

\begin{figure}
  \centering
  \begin{captivy}{\includegraphics[width=0.48\textwidth]{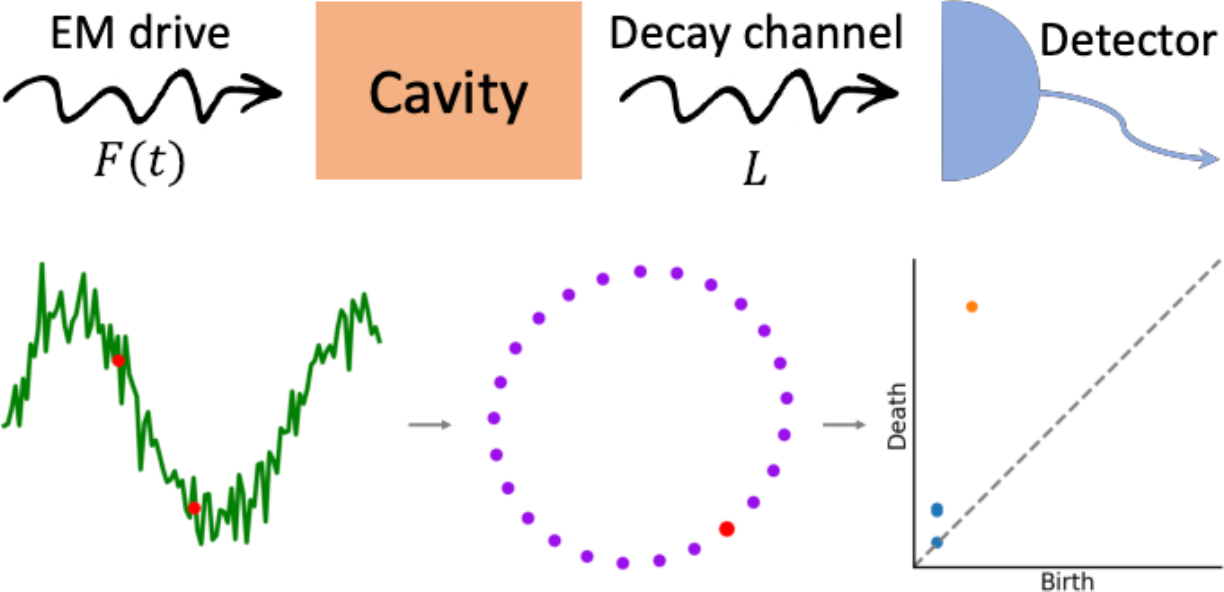}}%
  \oversubcaption{0, 0.9}{}{fig:schematic}
  \oversubcaption{0, 0.5}{}{fig:tda_pipeline}
  \end{captivy}
  \caption{Schematic of (a) leaky periodically-driven nonlinear cavity model and (b) TDA pipeline for a noisy periodic time series (left) embedded into two-dimensions using time-delay embedding (middle) with the presence of a single cycle indicated by the corresponding persistence diagram (right) with 0-D (blue) and 1-D (orange) features. To illustrate the embedding of two points (red) in the time series according to Takens' embedding, the corresponding two-dimensional point (red) is indicated on the point cloud.}
  \label{fig:illustration}
\end{figure}

We apply this approach to probe the structural stability of cavity-like open quantum systems in order to detect transitions to quantum chaos (that is, regimes where the classical counterpart is chaotic). Namely, we consider a Kerr-nonlinear cavity driven by a periodically modulated external EM field and show that the topological properties of a single quantum trajectory can be used to distinguish between regular and chaotic parameter regimes of the system. These quantum trajectories can be realised in continuous monitoring schemes, which opens a route for the application of this topological approach in real-life experimental setups \cite{Murch_2013, Weber_2016}. A graphical schematic of the full pipeline is shown in Fig. \ref{fig:illustration}. From a broader perspective, this work also bridges the application of TDA to quantum dynamics and we foresee that further research surrounding this connection will be fruitful.

The outline of this paper is as follows: Section II provides a brief introduction to attractor reconstruction using time-delay embedding and the TDA method of persistent homology. Section III establishes the quantum system (and its classical counterpart) considered in the current study and introduces the Monte Carlo wavefunction method. Sections IV and V present the proposed topological approach for identifying regular and chaotic regimes in classical and quantum systems respectively, demonstrating examples of the full pipeline. Section VI summarises the results obtained using this approach to detect chaotic phases in the parameter space of the system. Section VII makes concluding remarks and suggests future directions.
\section{\label{sec:background}Background}

\subsection{State space reconstruction}
\label{sec:reconstruction}
In general, the attractor of a classical dynamical system may not be directly accessible in an experimental setup because a mathematical description of the system and the dimensionality of its full phase space is unknown. Typically, measurements may only be able to be recorded of one relevant quantity at regular time intervals, resulting in a single time series $x(t)$ that represents a scalar observation function of the full state vector $\textbf{x}(t)$,
\begin{equation}
    x(t) = F(\textbf{x}(t)).
\end{equation}
Takens' embedding theorem implies that this single time series that has been observed from a potentially unknown dynamical system can be useful in reconstructing the attractor for the full system~\cite{takens_1981}. Specifically, the observed time series $x(t)$ can be `embedded' into a $d$-dimensional vector space in such a way as to preserve the topological properties of the \textit{full} state space $\textbf{x}(t)$. This \textit{time-delay embedding} involves taking uniformly time-lagged samples of the time series and concatenating them into a single vector, 
 
\begin{equation}
    \textbf{X}(t) = (x(t), x(t-\tau), ..., x(t-(d-1)\tau)),
\end{equation}

where $d$ is the embedding dimension and $\tau$ is the embedding time-delay. The embedded points form features, such as loops and voids, that reflect the same topology as the attractor of the dynamical system, guaranteed by Takens' theorem as long as appropriate embedding parameters are chosen \cite{takens_1981, ott_2002}. There is no consensus on the most favourable way to choose these and it is largely dependent on the specifics of the data, but popular heuristic methods include using mutual information for estimation of the optimal time-delay and false nearest neighbours for determination of a proper embedding dimension \cite{fraser_1986, kennel_1992}.

\subsection{Topological data analysis and persistent homology}
\label{sec:tda}

\begin{figure}
    \centering
    \includegraphics[width=0.48\textwidth]{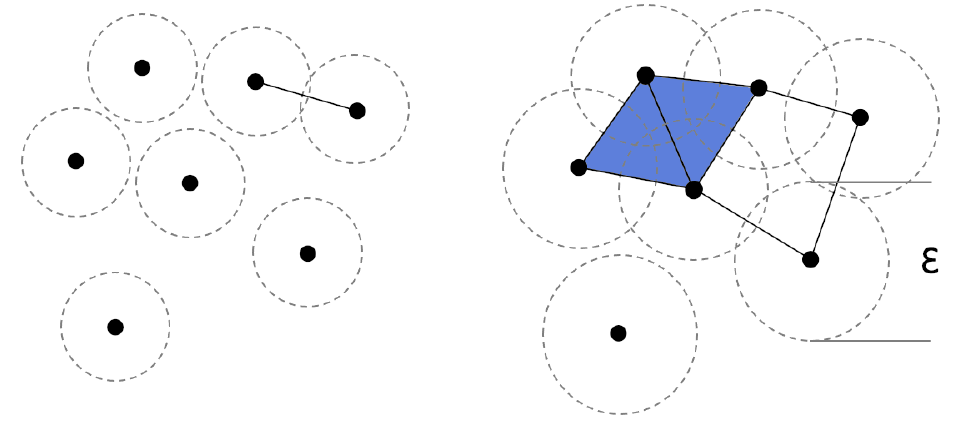}
    \caption{Simplicial complexes constructed at a small (left) and large (right) value of the filtration parameter $\epsilon$. Sets of $k$ points are connected together by $k-1$ simplices (edges, faces etc.) if all the points in the set have pairwise intersecting $\epsilon$-balls, corresponding to the Vietoris-Rips filtration.}
    \label{fig:complex}
\end{figure}

Topological data analysis (TDA) deals with attaching topological structure to finite sets of data, so that invariant properties under smooth deformation can be derived algebraically. These structures are known as \textit{simplicial complexes}, and can be constructed given any set of data that is equipped with some measure of distance. This triangulation-like procedure results in a network of \textit{k-simplices}, which are sets of $k+1$ of the points. For example, 0-simplices are points, 1-simplices are edges between pairs of points, 2-simplices are areas enclosed by three points, and so on. The condition under which $k$-simplices are formed from a finite set of points is chosen by the user, which consequently determines the k-simplices included in the simplicial complex. A common choice is the Vietoris-Rips (VR) complex which is built by forming $k$-simplices from sets of points whose pairwise distances are less than or equal to $\varepsilon$, a given non-negative filtration parameter (also referred to as a characteristic scale). The VR filtration is used in the present study and illustrated in Fig. \ref{fig:complex}. The reader should note that for much larger datasets, the exponential increase in the number of simplices makes computation extremely inefficient and other, more efficient, complexes can be used such as the witness complex, which builds the complex using a subset of the original data \cite{de_silva_2004}.

Over varying values of $\varepsilon$, a sequence of simplicial complexes can be obtained known as a \textit{filtration}. Each complex in the filtration will have different topological information which can be extracted to determine the abstract shape of that complex. One such non-trivial topological property is the number of $k$-dimensional ``holes", which can be determined by calculating the $k$-th Betti number of the complex using techniques from algebraic homology \cite{friedman_1998}. By tracking how the topology of the simplicial complex varies with the filtration parameter $\varepsilon$, we can effectively study what the shape of the dataset looks like at different scales. This method is known as $\textit{persistent homology}$, and allows a degree of importance to be attributed to topological features based on how far through the filtration they persist. The jargon refers to ranges of $\varepsilon$ as `lifetimes', with topological features being `born' or created at a particular value and `dying' or destroyed at another. Qualitatively, features that have short lifetimes we attribute to noise, while those that live for longer correspond to the `true' topological features of the dataset that we are aiming to discover.

In the above discussion we aim to introduce the method of persistent homology, although it is by no means a comprehensive introduction. We refer the reader to the following literature which provide a more in-depth and mathematical coverage, if this is needed \cite{zomorodian_2005, murugan_2019, leykam_2022}.

\section{\label{sec:model}Model}

For our quantum system of interest, we consider a damped, driven, Kerr-nonlinear cavity which is known to be chaotic in the classical case and whose chaotic regimes have been well studied in the literature \cite{szlachetka_1993, spiller_1994}. Its unitary dynamics is governed by the following Hamiltonian, written in a frame rotating with the cavity resonance frequency $\omega$,
\begin{equation}
 H = \frac{1}{2}\hbar\chi a^{\dagger2}a^{2}+i\hbar F(t)(a^{\dagger}-a),
 \label{hamiltonian}
\end{equation} where $a^{\dagger}$ ($a$) are photon creation (annihilation) operators respectively, $\chi$ is the photon interaction strength and $F(t)$ describes the periodic modulation by an external driving field. We consider the same driving as in Ref.~\cite{spiller_1994}, where $F(t)$ is a string of rectangular pulses of amplitude $A$, length $T/2$, and separation $T/2$, such that $F(t)=F(t+T)$. Note that if $F(t)$ is time-independent the system loses the potential for exhibiting chaotic behaviour.

Next, we incorporate damping of the cavity field into the model using the conventional formalism of open quantum systems. The time evolution of the reduced density matrix $\rho$ describing the cavity state is governed by a master equation of the Lindblad form \cite{BRE02},
\begin{equation}
    \frac{\partial \rho}{\partial t}=-\frac{i}{\hbar}[H,\rho]+\mathcal{L}[\rho].
    \label{lindbladian}
\end{equation}
The second term in the r.h.s. of Eq.~\eqref{lindbladian} captures the weak coupling of the system to the environment by a set of jump operators ${L_{i}}$. In general, there can be many of these operators which describe different classes of system-environment couplings. Ignoring this term simply reduces Eq.~\eqref{lindbladian} to the familiar von-Neumann equation, capturing the unitary dynamics of the system described by the Hamiltonian (\refeq{hamiltonian}). Here, only single photon emissions described by the jump operator $L$ are considered and the dissipative term in Eq.~\eqref{lindbladian} can be expressed as
\begin{equation}
\mathcal{L}[\rho]= L\rho L^{\dagger}-\frac{1}{2}\{L^{\dagger}L,\rho\},
\end{equation} where $\{A,B\}=AB+BA$. The effect of damping manifests as photon emissions from the leaky cavity, represented by a jump operator $L=\sqrt{\gamma}a$, where $\gamma$ is a dissipative coupling constant. We assume that the pumping rate of photons into the cavity from the thermal environment is zero, following Refs.~\cite{spiller_1994, yusipov_2020}.

Though the master Eq.~\eqref{lindbladian} can be solved numerically, if the relevant Hilbert space of the quantum system is of dimension $N$ the density operator requires $N^{2}$ real numbers to represent. This heavy computational requirement can be reduced using the quantum jump (Monte Carlo wavefunction) method, which constructs quantum trajectories by evolving wavefunctions $|\psi(t)\rangle$ using a pseudo-Hamiltonian of the form 
\begin{equation}
    H_{MC} = H - \frac{i\hbar}{2}L^{\dagger}L,
    \label{pseudo_hamiltonian}
\end{equation}
combined with stochastic jumps that correspond to the environmental action of the jump operator(s). Averaged over an ensemble of many trajectories, this evolution becomes equivalent to the master equation treatment in Eq.~\eqref{lindbladian} \cite{Molmer_1993, plenio_1998}. Expectation values of observables $\hat{O}$ are similarly calculated by using the expectation values of wavefunctions $\langle\psi(t)|\hat{O}|\psi(t)\rangle$ and taking the ensemble mean over all trajectories. Importantly, this method allows for mimicking the behaviour of individual realizations of the system dynamics and correctly reproduces the density matrix when averaged over many iterations.

\begin{figure}
    \centering
    \includegraphics[width=0.5\textwidth]{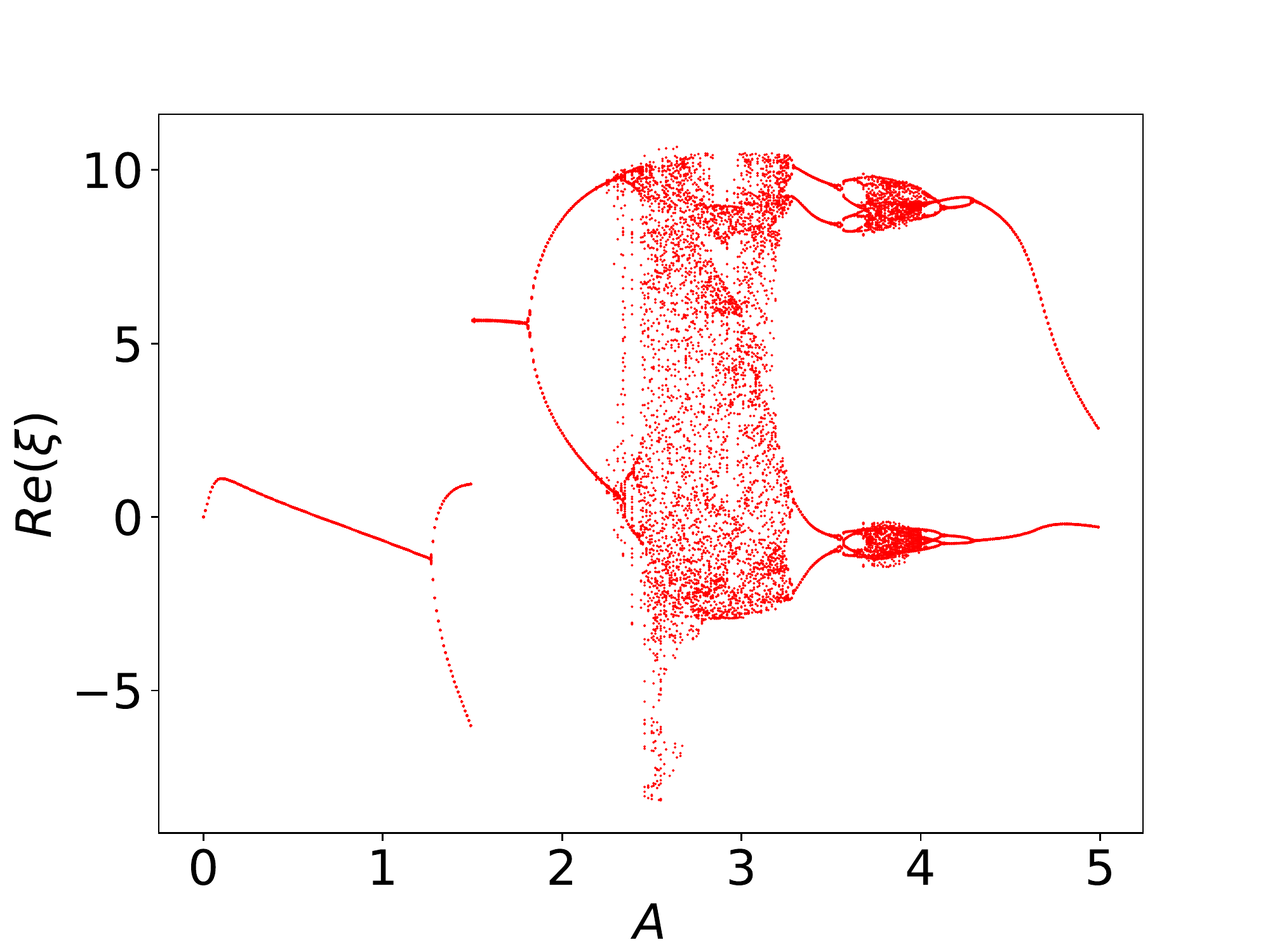}
    \caption{Stroboscopic values of $Re(\xi)$ at integer multiples $n$ of the driving period $40<n<100$ as a function of the driving amplitude $A$. This classical bifurcation diagram visually indicates the transitions between regular and chaotic behaviour. Simple curves correspond to regular periodic dynamics and the thicker bands correspond to chaotic trajectories which wander densely in the phase space. The driving period is fixed at $T=10$.}

    \label{fig:classical_bifurcation}
\end{figure}

In the classical limit of large amplitude coherent states, when the average number of photons in the cavity tends to infinity
, following Refs.~\cite{szlachetka_1993, spiller_1994, yusipov_2020} the master equation Eq.~\eqref{lindbladian} can be transformed into
\begin{equation}
    \frac{d\xi}{dt}=-\frac{1}{2}\gamma\xi+F(t)-i\chi|\xi|^{2}\xi^{*},
    \label{classical}
\end{equation} where $\xi$ is a complex dynamical variable whose real and imaginary components represent the position and momentum respectively. This variable can be compared directly with the expectation value of the photon annihilation operator $\langle\hat{a}\rangle_{\psi}$ in the quantum case. The nonlinear dynamical system described by Eq.~\eqref{classical} can be shown to exhibit regions of chaotic behaviour, as visualized in the bifurcation diagram [Fig.~\ref{fig:classical_bifurcation}] that shows the stroboscopic parameter dependence. Simple curves correspond to regular dynamics, whilst thicker bands indicate an exploration of the phase space that is characteristic of chaotic behaviour.

\section{Classical chaos detection}
\label{sec:classical_tda}

The attractor of a dissipative system describes its long-time evolution in phase space and has distinct topological properties based on whether the system dynamics exhibit regular (non-chaotic) or chaotic behaviour. As discussed in Sec. \ref{sec:reconstruction}, an attractor can be faithfully reconstructed (up to its topology) from the time-series of a single observable through an embedding technique, without knowledge of the full phase space dynamics. The resulting embedded data can then be analysed through TDA methods as persistent homology to characterise the reconstructed attractor and thus the system of interest. This method has been previously proposed as a quantifier for classical chaos, as an alternative to the maximal LE which can be difficult to compute for noisy, finite data \cite{maletic_2016, mittal_bifur_2017}.

Recall from the Introduction that in the study of quantum chaos we are interested in finding signatures of the quantum system in regimes where the corresponding classical system is chaotic. Thus, we start with an analysis of the classical limit described by Eq. \eqref{classical} and use it to determine the parameter regimes under which the system exhibits chaos.

\begin{figure}
  \centering
  \begin{captivy}{\includegraphics[width=0.5\textwidth]{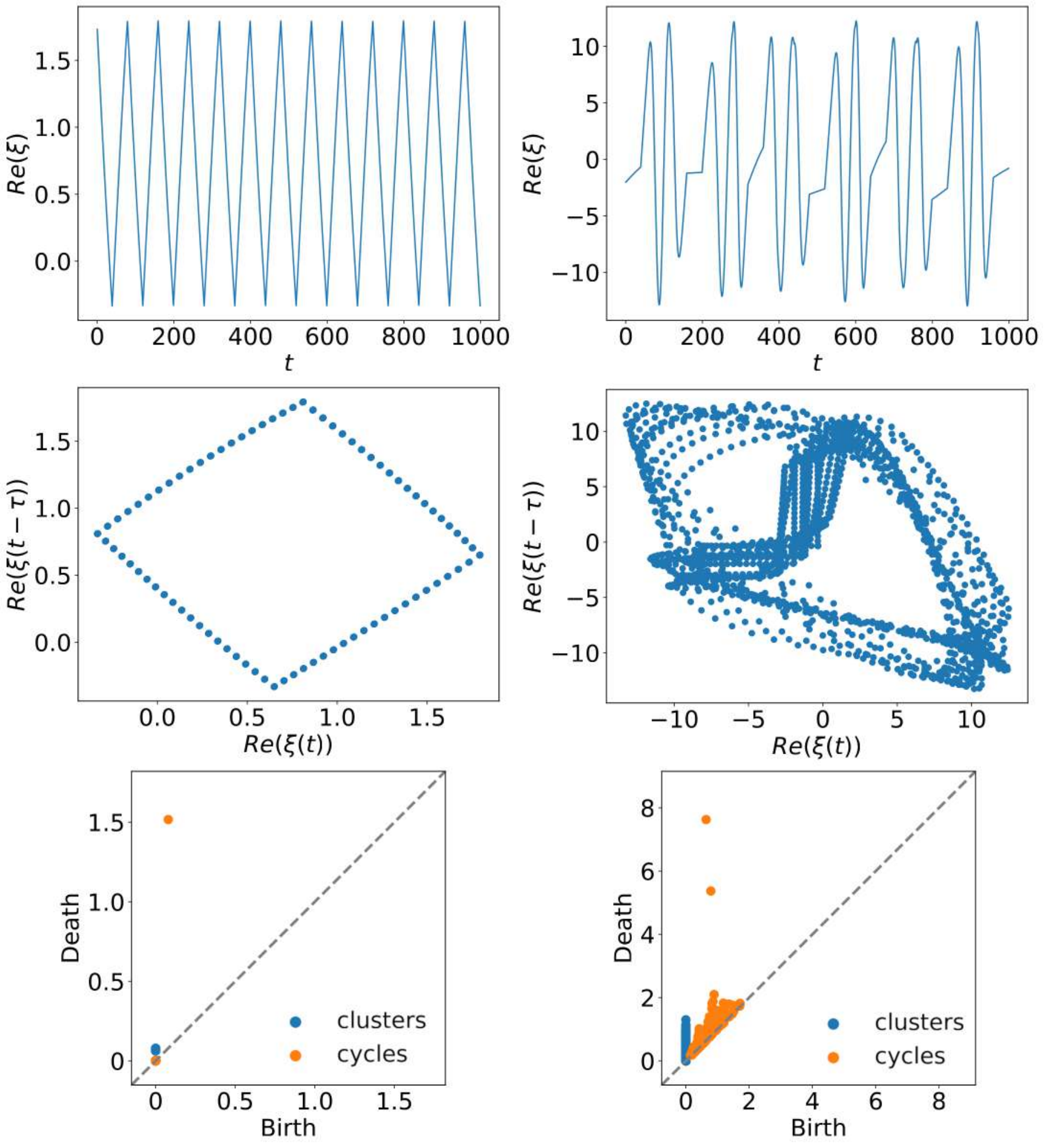}}%
  \oversubcaption{0, 0.95}{}{fig:classical_a}
  \oversubcaption{0.51, 0.95}{}{fig:classical_b}
  \oversubcaption{0, 0.63}{}{fig:classical_c}
  \oversubcaption{0.51, 0.63}{}{fig:classical_d}
  \oversubcaption{0, 0.3}{}{fig:classical_e}
  \oversubcaption{0.51, 0.3}{}{fig:classical_f}
  \end{captivy}
  \caption{Example of TDA pipeline for the classical system [Eq. \eqref{classical}] in regular ($A=1, T=8$) (left) and chaotic ($A=4.5, T=8$) (right) phases. Time series evolution of real part of dynamical variable (a,b) with corresponding two-dimensional time-delay embedding (c,d), and persistence diagram (e,f) for 0- (blue) and 1-dimensional (orange) features. Embedding time-delay is $\tau=20$, determined using heuristic methods (see Sec. \ref{sec:reconstruction}).}
  \label{fig:classical_pipeline}
\end{figure}

The embedding and TDA pipeline is demonstrated for the analysis of the classical dynamical system for drive parameter values corresponding to regular and chaotic phases in Fig. \ref{fig:classical_pipeline}. In this work, for the time-delay embeddings we determine optimal hyperparameters using the heuristic methods discussed in Sec. II. It should be noted that although the specific values of the embedding delay and embedding dimension are not important as long as they satisfy the criteria of Takens’ theorem~\cite{takens_1981}, they do have an influence on the features obtained through persistent homology due to the discrete nature of the data. We verify in the Appendix that the results for the quantum case remain robust over a wide range of hyperparameter values.

For driving parameters chosen in the regular phase ($A = 1$, $T = 8$), the periodic dynamics obtained by numerically solving Eq.~\eqref{classical} and shown in Fig.~\ref{fig:classical_pipeline}(a) lead to a reconstructed attractor that has a simple geometric structure. The embedded point cloud shown in Fig.~\ref{fig:classical_pipeline}(c) exhibits a single cycle. This is quantified by the persistence diagram in Fig.~\ref{fig:classical_pipeline}(e) which identifies the presence of the cycle through the single long-lived one-dimensional feature furthest from the diagonal. The strange attractors that correspond to chaotic dynamics have much more complex structures whose topology cannot be described analytically as in the regular case \cite{ott_2002}. Fig. \ref{fig:classical_pipeline}(b) shows the time series for parameter values chosen in the chaotic phase ($A = 4.5$, $T = 8$), which results in a reconstructed attractor whose points are scattered densely over the phase space in Fig.~\ref{fig:classical_pipeline}(d). Correspondingly, the persistence diagram in Fig.~\ref{fig:classical_pipeline}(f) indicates that at the scale of the system dynamics there are numerous features with a range of lifetimes. 
During the transition between the regular and chaotic phases, we note that the attractors develop two long-lived features indicative of two cycles. This is highly suggestive of period-doubling behaviour on the route to chaos, also illustrated by the curve bifurcations in Fig. \ref{fig:classical_bifurcation}, as the correlation between the number of cycles observed in phase space trajectories and multi-periodicity of the dynamics has been previously identified \cite{mittal_bifur_2017, linsay_1981}.

These distinctions between persistent features of different systems can be used as a quantifier for distinguishing regular and chaotic phases. Motivated by the significance of the size and number of cycles outlined in the above discussion, we consider as topological measure the average lifetime of low-dimensional topological features, discussed further in Sec. VI.

\section{Quantum chaos detection}
\label{sec:quantum_tda}

Turning now to the quantum case, we present an approach for detecting chaotic dynamics in open quantum systems using TDA, which draws inspiration from the classical attractor reconstruction method outlined above. The key idea is that instead of the ensemble-based master equation formulation, single quantum trajectories obtained from the quantum jump unravelling can be used as direct analogues to the trajectories in the classical formulation of chaos. Through measurement of a suitable observable, a time series can be obtained whose dynamical properties we can extract through time-delay embedding and the application of TDA as described in Sec. \ref{sec:classical_tda}.

It has been shown previously that as an open quantum system becomes macroscopic, the behaviour of observables $\langle\hat{x}\rangle$ and $\langle\hat{p}\rangle$ gradually resemble that of classical attractors in phase space \cite{Brun_1996, ota_2005}. However, at small action scales deep in the quantum regime, this classical evolution becomes blurred by decoherence and stochastic dissipative influences of the environment. Our findings suggest that the stochastic evolution of individual states is able to reflect the topology of the underlying motion, robust to the effects of quantum uncertainty. As a consequence, topological analysis of single quantum trajectories provides a quantifier for quantum chaos in open quantum systems. Without requiring the dynamics of the entire density operator, this method is computationally efficient and, more importantly, can be easily applied in experimental settings. For example, single quantum trajectories can be realised through continuous monitoring schemes such as homodyne detection, or through directly measurable observables such as single photon detections (or any other decay/pump channel) corresponding to quantum jumps~\cite{Murch_2013, Weber_2016}.

\begin{figure}
  \centering
  \begin{captivy}{\includegraphics[width=0.5\textwidth]{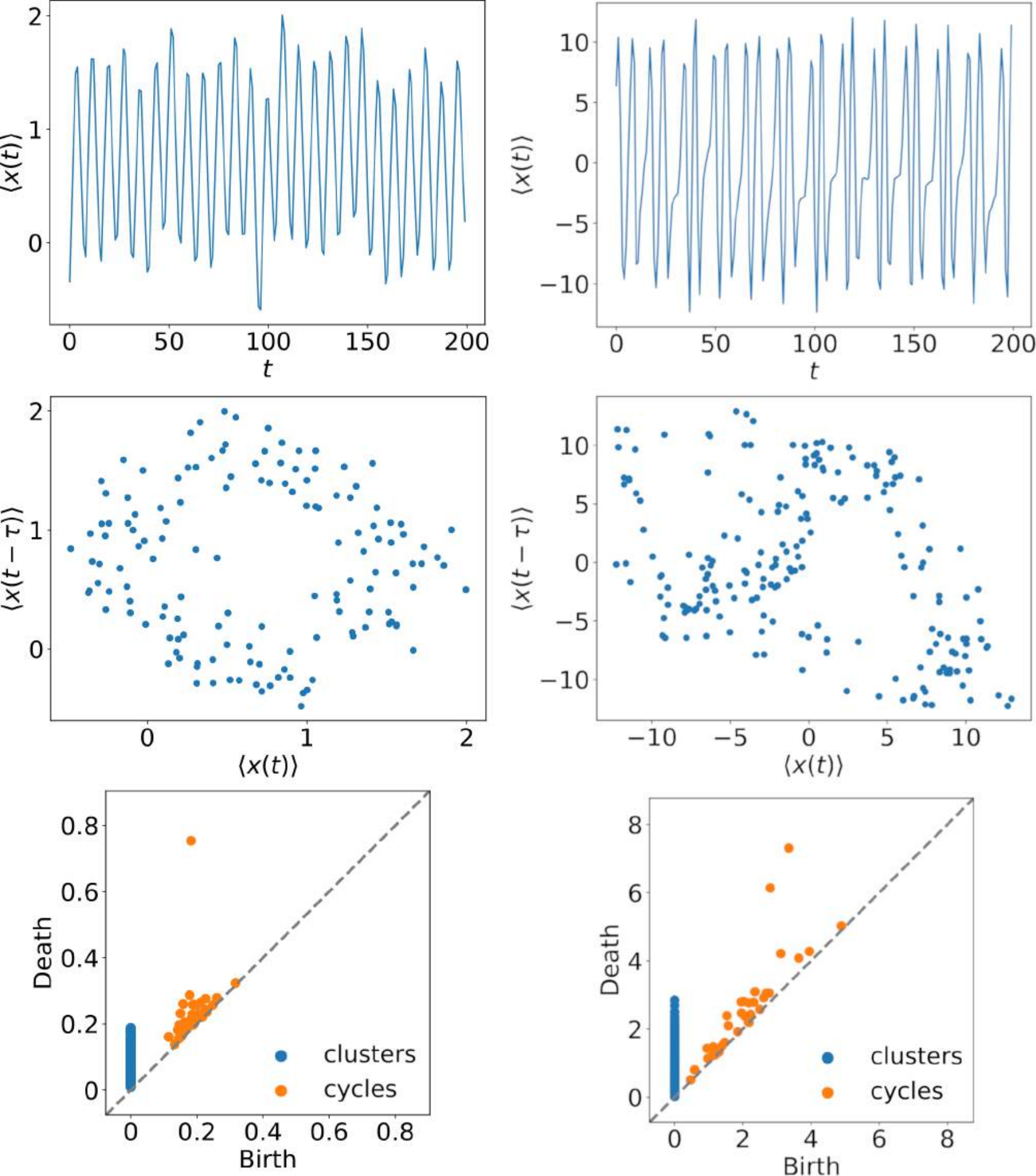}}%
  \oversubcaption{-0.01, 0.965}{}{fig:quantum_a}
  \oversubcaption{0.51, 0.965}{}{fig:quantum_b}
  \oversubcaption{-0.01, 0.63}{}{fig:quantum_c}
  \oversubcaption{0.51, 0.63}{}{fig:quantum_d}
  \oversubcaption{-0.01, 0.295}{}{fig:quantum_e}
  \oversubcaption{0.51, 0.295}{}{fig:quantum_f}
  \end{captivy}
  \caption{Example of TDA pipeline for the quantum system [Eq. \eqref{hamiltonian}] in regular ($A=1, T=8$) (left) and chaotic ($A=4.5, T=8$) (right) phases. Time series evolution of position expectation value $\langle x\rangle$ (a,b) with corresponding two-dimensional time-delay embedding (c,d), and persistence diagram (e,f) for 0- (blue) and 1-dimensional (orange) features. Embedding time-delays are $\tau=6$ and $\tau=2$ respectively, determined using heuristic methods (see Sec. \ref{sec:reconstruction}). Note that there are fewer points in the embedded cloud compared to the corresponding classical clouds due to the shorter evolution time simulated.)}
  \label{fig:quantum_pipeline}
\end{figure}

We first demonstrate this approach by considering the time evolution of the position expectation value $\langle\hat{x}\rangle$ as our observable to provide a simple comparison with the classical case in Fig.~\ref{fig:quantum_pipeline}. The time-evolution of the quantum system described by Eq. \eqref{hamiltonian} is simulated using the computational library \textit{QuTiP} \cite{qutip}. We allow a single quantum trajectory to evolve over $T_{trans}<t<1000T$, where $T_{trans} = 400T$ is discarded as transient behaviour. We use Hamiltonian parameters $\chi=0.008$ and $\gamma=0.05$, which gives an average photon number $~50$, and consider a truncated Hilbert space with up to $N=300$ photons. Using the computed wavefunctions which evolve stochastically under Eq. \eqref{pseudo_hamiltonian}, we first obtain the time-evolution for the position expectation value $\langle \hat{x} \rangle$ and apply time-delay embedding as before to obtain a higher-dimensional point cloud that captures an underlying topology of the system dynamics through Takens' theorem. Since the point clouds in Fig.~\ref{fig:quantum_pipeline}(c,d) are observed qualitatively to resemble the attractors of the dissipative classical case, we refer to them as `quantum attractors'. In the classical limit with large mean photon number, the single photon decays do not appreciably affect the trajectories and the classical results in Fig.~\ref{fig:classical_pipeline} are reproduced exactly. For the moderate photon numbers considered here, there is an obvious smearing out of the point clouds due to quantum noise arising from fluctuations in the photon number from stochastic decay events. 

Despite the presence of this noise in the quantum regime, the key result to note here is that the topological features of these noisy quantum attractors remain robust. As a consequence, these features can be extracted using persistent homology and used to differentiate dynamical phases as in the classical case discussed in Sec. \ref{sec:classical_tda}. Here, we focus on low-dimensional features which are more efficient to compute than high-dimensional features and capture  more intuitively meaningful properties of the topology. We expect, however, that higher dimensional topological features may provide additional useful insights and leave this study to future work.

Given that a true topological description of the dynamics has been captured, we expect the topology extracted by Takens' theorem to be independent of the observable chosen. Thus, we also consider an observable based on the count of quantum jump events, which is easier to obtain experimentally and does not interfere with the intra-cavity dynamics. A time series of binned photon counts is constructed by moving a sliding window across the quantum jump times, as shown in Fig. \ref{fig:photon_count}(a), which we then process using time-delay embedding in a similar manner. The topological summary results for both observables will be further discussed in Sec. \ref{sec:topological_summary}.

\section{Topological summaries}
\label{sec:topological_summary}

Various summaries can be used for quantifying the topological information of persistence diagrams, chosen depending on the specific use-case \cite{ leykam_photonic_2021, mittal_bifur_2017, guzel_2022}. Here, we consider the average lifetime of low-dimensional features in the persistence diagram, defined as
\begin{equation}
L_{avg}(D) = \frac{1}{|D|}\sum_{h\in D}(d - b),
\end{equation} where $D$ is the set of all features in the persistence diagram, and $b$ ($d$) corresponds to the $\epsilon$-value of birth (death) of a particular feature $h$. This provides a summary similar to that of the Betti number and tells us the average size of topological feature in the reconstructed attractor. 

\begin{figure}
  \centering
  \begin{captivy}{\includegraphics[width=0.5\textwidth]{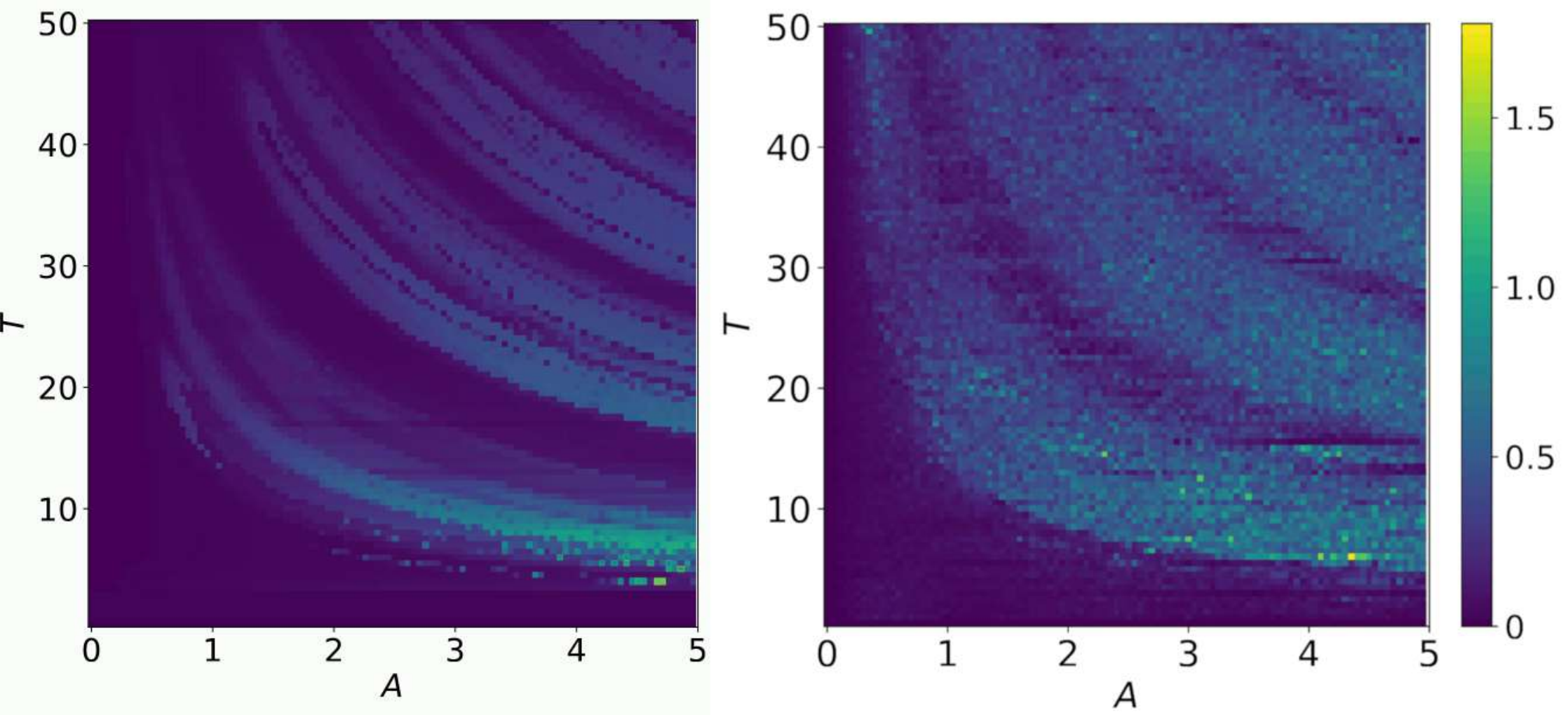}}%
  \oversubcaption{0, 0.9}{}{fig:classical_phase}
  \oversubcaption{0.47, 0.9}{}{fig:quantum_phase}
  \end{captivy}

  \caption{Average 1-hole lifetime as a function of driving amplitude $A$ and driving period $T$ for time evolution of (a) real part of dynamical variable for the classical system [Eq. \eqref{classical}] and (b) position expectation value $\langle x \rangle$ of single Monte Carlo trajectory of quantum system [Eq. \eqref{hamiltonian}].}
  \label{fig:full_params_1}
\end{figure}

It is known that under different drive parameters $A$ and $T$, the quantum system of Eq. \eqref{hamiltonian} contains regions of regular and chaotic behaviour \cite{yusipov_2019, yusipov_2020}. Here, we consider the driving amplitude between the range of $0<A<5$ and driving period between $0<T<50$. The average lifetime for 1-dimensional holes of the reconstructed attractors are computed over this ($A$,$T$)-parameter space for both the classical and quantum systems, shown in Figs.~\ref{fig:full_params_1} (a) and (b), respectively. The key observation is that the distinct chaotic bands defined by different topological properties of the corresponding quantum attractors are well correlated with the dynamical phases captured by the TDA-based approach applied to the classical system. This suggests that the quantum attractors reconstructed from single quantum trajectories contain topological information about the open quantum system that can be used to distinguish between regular and chaotic dynamics. Note that the blurring of fine structure variations of the classical dynamics within the chaotic phase bands of Fig. \ref{fig:full_params_1}(a) are due to effects arising from the quantum-classical transition \cite{pokharel_2018, yusipov_2020}. Additionally, we show that this measure is a robust indicator of chaos by demonstrating that the phase distinguishability is independent of free hyperparameters for the time-delay embedding as long as they satisfy the conditions for Takens' theorem (see Appendix).

\begin{figure}
  \centering
  \begin{captivy}{\includegraphics[width=0.5\textwidth]{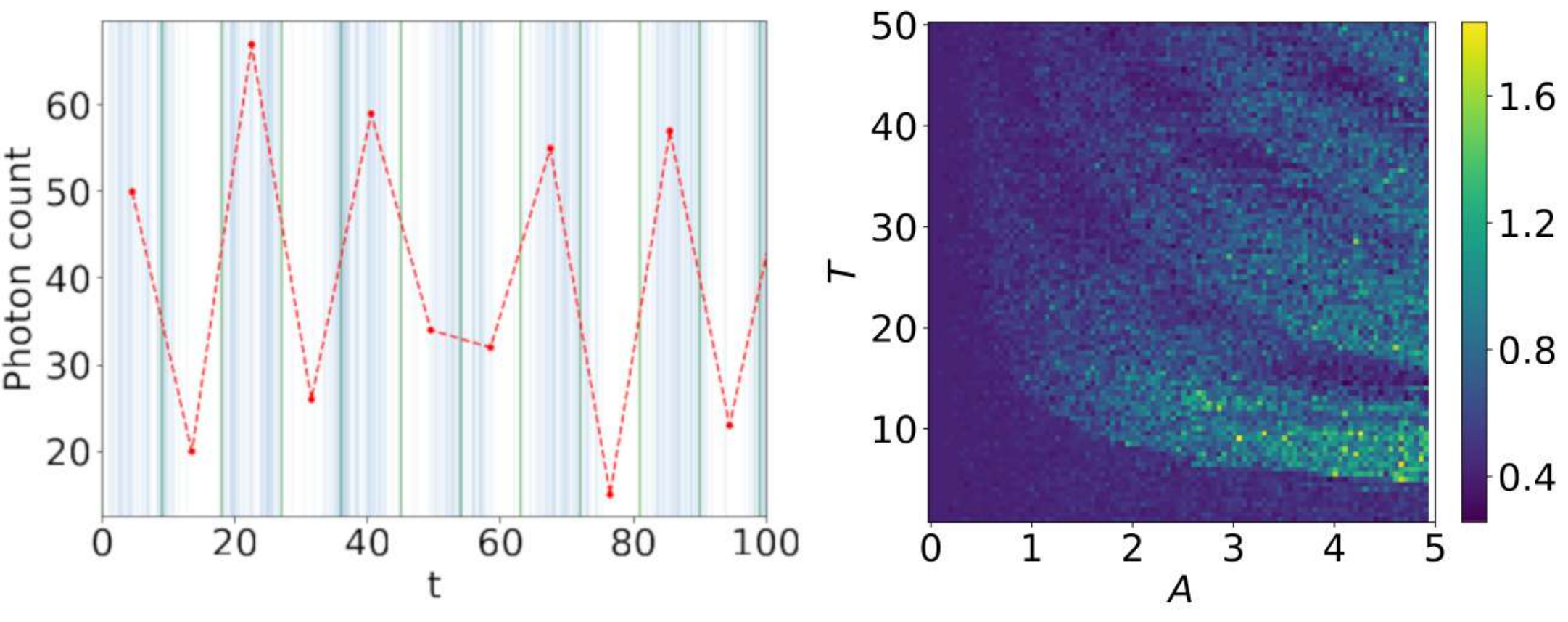}}%
  \oversubcaption{0, 0.88}{}{fig:photon_count_example}
  \oversubcaption{0.52, 0.88}{}{fig:photon_phase}
  \end{captivy}

  \caption{(a) Binned photon count time series (red dashed line) for chaotic phase ($A=4.5$, $T=8$) with bin edges (green) and jump event timings (blue). (b) Average 1-hole lifetime as a function of driving amplitude $A$ and driving period $T$. The binning interval is fixed at $9$ and embedding parameters chosen using heuristic methods (see Sec. \ref{sec:reconstruction}).}
  \label{fig:photon_count}
\end{figure}

Figure~\ref{fig:photon_count}(b) shows the corresponding phase diagram obtained by considering the photon count as the observable and is similarly reminiscent of the classical phase diagram in Fig. \ref{fig:full_params_1}(a). It should be noted that the additional free parameter of the binning interval size has a large influence on these values and the distinction between phases is somewhat less clear. The intervals need to be sufficiently large to filter out zero photon count bins, and consequently the time step-size is larger than for the position operator observable, which results in fewer embedded points for a fixed evolution time. Nonetheless, we emphasize the observation that this simple measurable offers insight into a topological signature of quantum chaos and we leave it to future studies to investigate the optimal observables for probing such topologies of quantum dynamics.
\section{\label{sec:conclusion}Conclusion}

In this work we introduced a topological approach for the dynamical analysis of open quantum systems, which applies a phase-space embedding method on single quantum trajectories to construct `quantum attractors' and extract their topological properties using persistent homology. This was inspired by recent applications of topological data analysis to the dynamical analysis of classical systems, underpinned by Takens' theorem which guarantees the preservation of topology under certain transformations of time series data. 

We used our quantum pipeline to investigate the dynamical phases of a leaky Kerr-nonlinear photonic cavity driven by an external field, which can exhibit chaotic behaviour under certain driving parameters. Our results indicate that the single quantum trajectory retains topological information about the entire system dynamics which can be extracted to capture the regular and chaotic phases of the open quantum system. Importantly, these topological measures are found not to be dependent on the observable considered and can be calculated using limited measurement of the system through weak continuous monitoring schemes or other directly measurable observables. Thus, this approach presents an advantage over current methods for quantifying chaos in open quantum systems which are experimentally difficult to realize \cite{Swingle_2016,Li_2017,macri_2016, Zanardi_2021}. 

To the best of our knowledge, this work is the first to apply the framework of topological data analysis for the analysis of chaotic quantum dynamics. We hope that this can open up a new perspective for employing such topological machine learning techniques for studying the dynamics of quantum systems, making use of their inherent robustness to noise and limited measurement and ability to effectively handle high-dimensional datasets.

Since the TDA-based approach can be applied to both the classical and quantum regimes, a possible future route of investigation would be to see whether these topological descriptions of quantum systems can offer any insight into critical points during the quantum-classical transition. For instance, the TDA method of zig-zag persistence may provide a convenient way of tracking the persistence of specific features through a filtration of quantum attractors that correspond to changing size/action scales of the dynamical system \cite{carlsson_2008, Tymochko_2020}.

Another direction for which this TDA approach may aid in uncovering new physics is understanding quantum phase transitions in relation to the time dependence of entanglement structures, following recent work using persistent homology to extract experimentally-intractable entanglement measures of quantum many-body systems \cite{piga_2019, Zanardi_2021, boixo_2018}.

Finally, we note the considerable attention given to quantum TDA algorithms as a promising candidate for near-term quantum advantage \cite{gyurik_2020, akhalwaya_2022}. In particular relation to our work, a recent paper \cite{ameneyro_2022} establishes a quantum Takens' embedding algorithm that may expedite a search for a fully quantum pipeline for the study of quantum chaos.

\appendix*

\section{Time-delay embedding parameter dependence}
\label{sec:appendix}

It should be noted that although the specific values of the time delay and dimension for time-delay embedding are not important as long as they satisfy the criteria of Takens' theorem \cite{takens_1981}, they do have an influence on the features obtained through persistent homology due to the discrete nature of the data. 

\begin{figure}
  \centering
  \begin{captivy}{\includegraphics[width=0.4\textwidth]{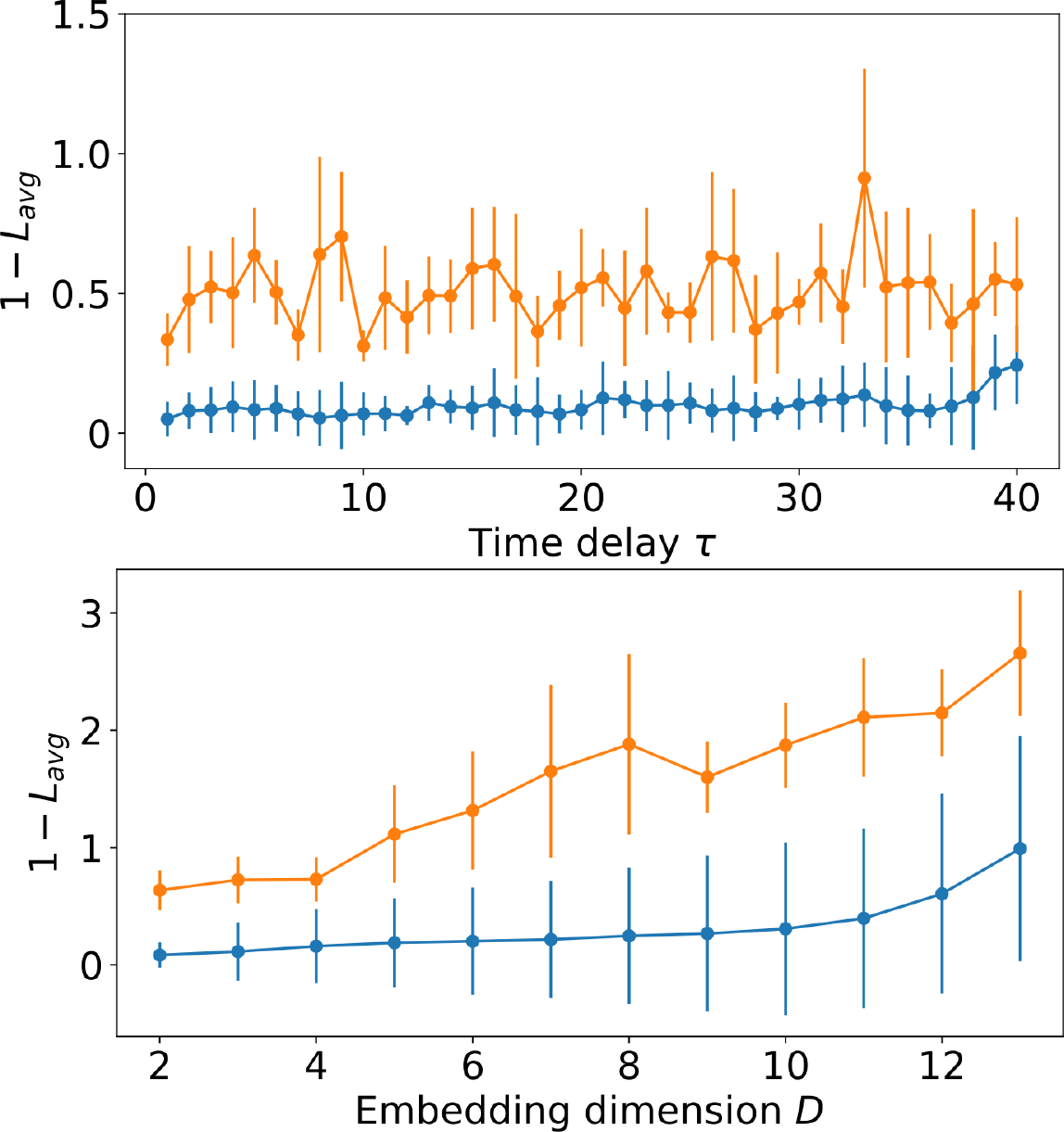}}%
  \oversubcaption{0, 0.95}{}{fig:time_delay}
  \oversubcaption{0, 0.45}{}{fig:dimension}
  \end{captivy}

  \caption{Average 1-hole lifetime as a function of (a) embedding time delay ($D=2$) and (b) embedding dimension ($\tau=7$) for position expectation value $\langle x \rangle$ of single Monte Carlo trajectory of quantum system [Eq. (3)], averaged over 10 points in the regular (blue) and chaotic (orange) phases respectively with corresponding standard deviations.}
  \label{fig:parameters}
\end{figure}

Here, we demonstrate that the results for the quantum case remain robust over a range of hyperparameter values. Figure \ref{fig:parameters}(a) shows that for a fixed embedding dimension and varying the time-delay parameter, the topological measure averaged over 10 different chosen points in the regular and chaotic phases respectively can still be easily distinguished. Figure \ref{fig:parameters}(b) indicates the same result, but with a fixed time-delay and varying the embedding dimension.

\bibliography{references}

\end{document}